\documentclass[aip,reprint]{revtex4-1}

\usepackage{graphicx}
\usepackage{dcolumn}
\usepackage{bm}
\usepackage{textcomp}
\usepackage{gensymb}
\usepackage[articletitle=true]{achemso}
\usepackage[utf8]{inputenc}

\draft 

\DeclareUnicodeCharacter{2212}{-}

\begin{document}

\preprint{AIP/123-QED}

\title{A Reversible Structural Phase Transition by Electrochemical Ion Injection into a Conjugated Polymer}

\author{Connor G. Bischak}
\affiliation{Department of Chemistry, University of Washington, Seattle, Washington 98195-1700, United States}

\author{Lucas Q. Flagg}
\affiliation{Department of Chemistry, University of Washington, Seattle, Washington 98195-1700, United States}

\author{Kangrong Yan}
\affiliation{MOE Key Laboratory of Macromolecular Synthesis and Functionalization, State Key Laboratory of Silicon Materials, Department of Polymer Science and Engineering, Zhejiang University, Hangzhou 310027, P.R. China}

\author{Kangrong Yan}
\affiliation{MOE Key Laboratory of Macromolecular Synthesis and Functionalization, State Key Laboratory of Silicon Materials, Department of Polymer Science and Engineering, Zhejiang University, Hangzhou 310027, P.R. China}

\author{Tahir Rehman}
\affiliation{Department of Chemical and Biomolecular Engineering, University of Illinois at Urbana−Champaign, 600 South Mathews Avenue, Urbana, Illinois 61801, United States}

\author{Ramsess J. Quezada}
\affiliation{Department of Chemistry, University of Washington, Seattle, Washington 98195-1700, United States}

\author{Jonathan W. Onorato}
\affiliation{Department of Materials Science and Engineering, University of Washington, Seattle, Washington 98195, United States}

\author{Christine K. Luscombe}
\affiliation{Department of Materials Science and Engineering, University of Washington, Seattle, Washington 98195, United States}
\affiliation{Department of Molecular Engineering and Sciences, University of Washington, Seattle, Washington 98195, United States}

\author{Ying Diao}
\affiliation{Department of Chemical and Biomolecular Engineering, University of Illinois at Urbana−Champaign, 600 South Mathews Avenue, Urbana, Illinois 61801, United States}

\author{Chang-Zhi Li}
\affiliation{MOE Key Laboratory of Macromolecular Synthesis and Functionalization, State Key Laboratory of Silicon Materials, Department of Polymer Science and Engineering, Zhejiang University, Hangzhou 310027, P.R. China}

\author{David S. Ginger}
\altaffiliation{Corresponding Author (dginger@uw.edu)}
\affiliation{Department of Chemistry, University of Washington, Seattle, Washington 98195-1700, United States}

\begin{abstract}
We find that conjugated polymers can undergo reversible structural phase transitions during electrochemical oxidation and ion injection. We study poly[2,5-bis(thiophenyl)-1,4-bis(2-(2-(2-methoxyethoxy)ethoxy)ethoxy)benzene] (PB2T-TEG), a conjugated polymer with glycolated side chains.  Using grazing incidence wide angle X-ray scattering (GIWAXS), we show that, in contrast to previously known polymers, this polymer switches between two structurally distinct crystalline phases associated with electrochemical oxidation/reduction in an aqueous electrolyte. Importantly, we show that this unique phase change behavior has important physical consequences for ion transport. Notably, using moving front experiments visualized by both optical microscopy and super-resolution photoinduced force microscopy (PiFM), we show that a propagating ion front in PB2T-TEG exhibits non-Fickian transport, retaining a sharp step-edge profile, in stark contrast to the Fickian diffusion more commonly observed. This structural phase transition is reminiscent of those accompanying ion uptake in inorganic materials like LiFePO\textsubscript{4}. We propose that engineering similar properties in future conjugated polymers may enable the realization of new materials with superior performance in electrochemical energy storage or neuromorphic memory applications.
\end{abstract}


\maketitle 
The movement of ions within materials, and the transport of ions across material interfaces are fundamental processes that underlie many technologies, including bioelectronics,\cite{Someya2016rise,Simon2016Organic} mechanical actuators,\cite{Jager2000Microfabricating,Lu2002Use} light emission,\cite{Pei1995Polymer} neuromorphic computing,\cite{Burgt2017non-volatile,Fuller2019Parallel} and including energy storage.\cite{Tang2010Electrochemically,Nitta2015Li-ion} Electrochemical ion insertion is often accompanied by a reorganization of the host material, with the host expanding to accommodate the volume of the additional ions. In inorganic materials, such as lithium ion battery electrodes, ion insertion often leads to a structural phase transition of the host material, in which the presence of ions induces a new thermodynamically-favorable structural phase.\cite{Padhi1997Phosphoolivines,Lim2016Origin} A consequence of the structural phase transition is phase separation between ion-rich and ion-poor regions of the host material at particular ion concentrations. In lithium ion battery electrode materials, this phase separation results in a flat voltage profile upon battery discharging.\cite{Padhi1997Phosphoolivines} On the other hand, soft materials, such as conjugated polymers or polymer electrolytes, which are often semi- or paracrystalline, generally deform without undergoing a structural phase transition upon ion insertion, and phase separation between the ion-rich and ion-poor regions is not observed.\cite{Smela2001Volume,Hallinan2013Polymer,Guardado2017Structural,Thomas2018X-Ray} 

Conjugated polymers are an important class of organic electronic materials that are capable of transporting both electronic and ionic charge carriers.\cite{Paulsen2019Organic} Compared to inorganic materials, such as silicon, conjugated polymer have lower electronic mobilities, yet are capable of fast, reversible, volumetric charge injection.\cite{Paulsen2019Organic,Inal2018Conjugated} Recently, the ability of these mixed conductors to uptake and transport both ionic and electronic charges has led to their study in applications including bioelectronics,\cite{Berggren2007Organic,Rivnay2014Rise,Rivnay2018Organic} thermoelectrics,\cite{Bubnova2012Towards} neuromorphic computing,\cite{Burgt2017non-volatile} and energy storage.\cite{Mike2013Recent,Bryan2016Conducting,Moia2019Design} Many of these technologies rely on ion injection through the application of an electrochemical bias, which oxidizes (or reduces) the polymer, driving anions (or cations) into the conjugated polymer where they provide charge compensation for the injected electronic carriers, allowing the realization of capacitances as high as ~1000 F/g.\cite{Bryan2016Conducting} Although this process provides the foundation for many technologies, the mechanism of ion injection and transport in mixed conductors is still an active area of research.\cite{Someya2016rise,Simon2016Organic}

\begin{figure*}
\includegraphics[width=10cm]{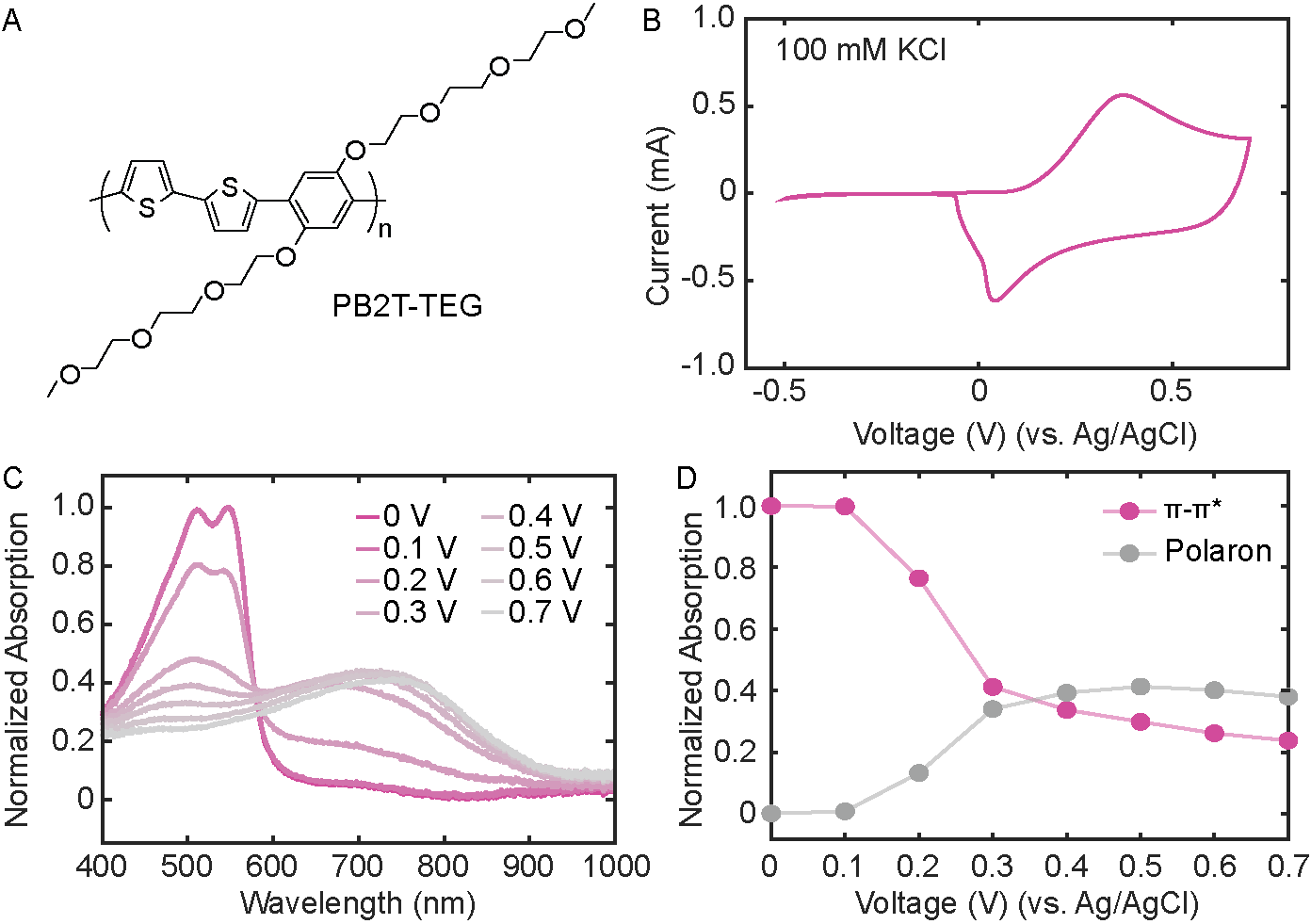}
\caption{\label{fig1} Electrochemical and spectroelectrochemistry characterization of PB2T-TEG. (A) Chemical structure of PB2T-TEG. (B) Cyclic voltammetry (CV) of PB2T-TEG in 100 mM KCl at a rate of 0.1 V/s. (C) Absorption spectra as a function of applied voltage (vs. Ag/AgCl), showing a decrease in the $\pi$-$\pi$* absorption peak and an increase of the polaron absorption peak. (D) Normalized absorption versus applied voltage (vs. Ag/AgCl) for the $\pi$-$\pi$* absorption peak (512 nm) and the polaron absorption (720 nm). }
\end{figure*}

The incorporation of molecular dopants and ions in conjugated polymers alters the electronic conductivity of the conjugated polymer host and impacts its crystal structure.\cite{Kroon2016Thermoelectric,Jacobs2017Controlling} The structural changes that conjugated polymers undergo upon electrochemical ion insertion have been investigated in a number of cases.\cite{Guardado2017Structural,Thomas2018X-Ray,Flagg2019Polymer} For instance, in poly(3-hexylthiophene-2,5-diyl) (P3HT), the lamellar spacing of the polymer increases continuously and the $\pi$-$\pi$ stacking distance decreases as the concentration of injected ions increases.\cite{Guardado2017Structural,Thomas2018X-Ray} Although structural changes in P3HT have been widely studied, P3HT is a relatively poor conjugated polymer for transporting most ions due to its hydrophobic alkyl side chains.\cite{Flagg2018Anion-Dependent} Recently, new conjugated polymers and small molecule semiconductors designed for enhanced ion transport with ethylene glycol or charged side chains have been synthesized.\cite{Flagg2019Polymer,Giovannitti2016Controlling,Nielsen2016Molecular,Inal2017Benchmarking,Moser2019Materials,Bischak2019Fullerene}, It is unclear whether these polymers behave similarly to P3HT or undergo different mechanisms for ion injection and reorganization of the polymer matrix, particularly under conditions relevant to device operation. 

Here, we investigate structural changes in poly[2,5-bis(thiophenyl)-1,4-bis(2-(2-(2-methoxyethoxy)ethoxy)ethoxy)benzene] (PB2T-TEG), a conjugated polymer with glycolated side chains, upon electrochemical oxidation using both ex situ and in operando grazing incidence X-ray scattering (GIWAXS). We find that, unlike other polymers studied to date, PB2T-TEG undergoes a reversible structural phase transition upon oxidation in an aqueous electrolyte. Using moving front experiments monitored by both optical transmission and nanoscale infrared imaging with photoinduced force microscopy (PiFM),\cite{Kong2018Identifying,Flagg2019Polymer} we observe a sharp ion front, inconsistent with Fickian diffusion of ions. We propose that structural phase transitions in conjugated polymers may be beneficial for many applications ranging from energy storage and to post-von Neumann computing.

\begin{figure*}
\includegraphics[width=16cm]{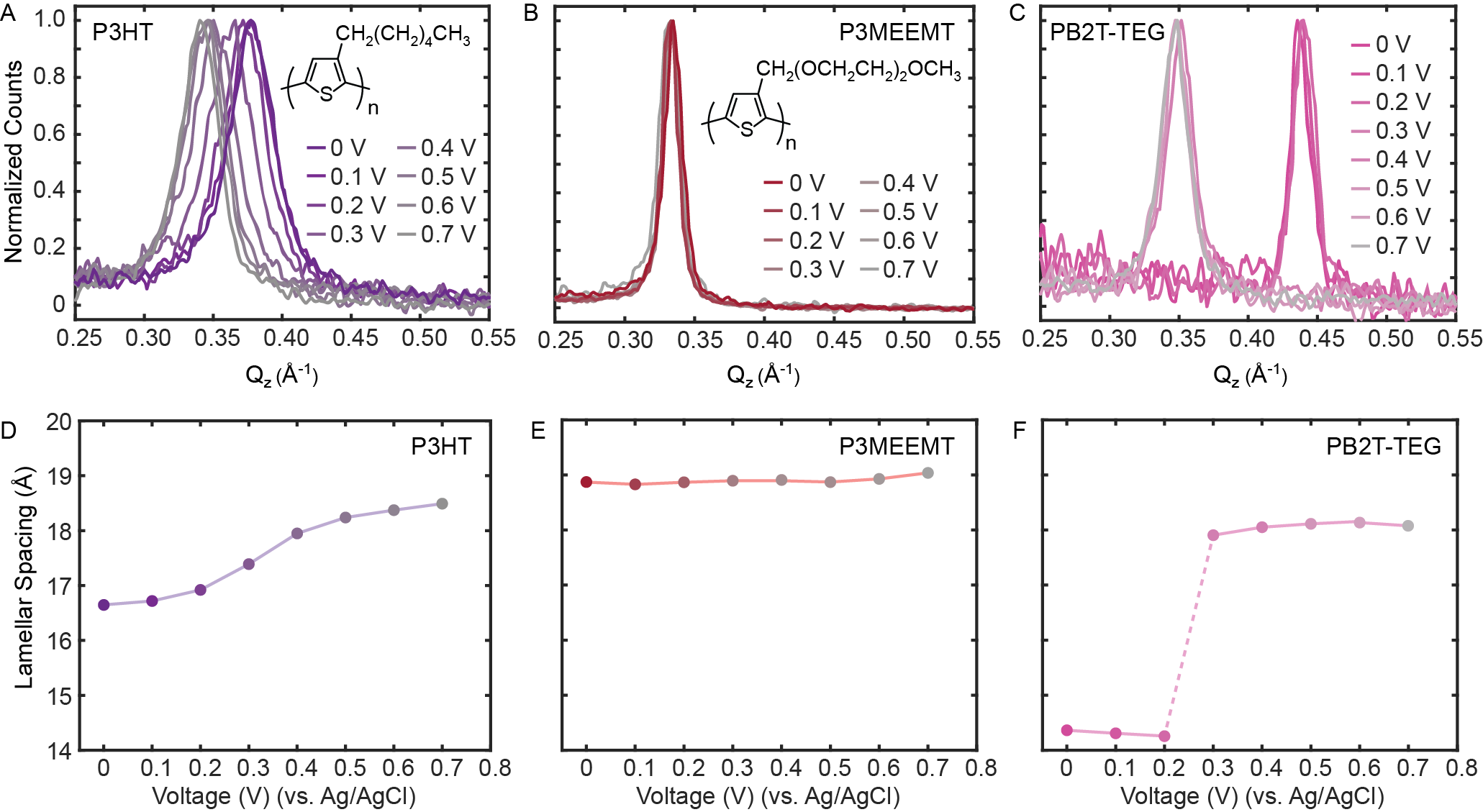}
\caption{\label{fig2} Structural changes upon electrochemical ion insertion for P3HT, P3MEEMT, and PB2T-TEG. XRD spectra of the (100) lamellar spacing diffraction peak for (A) P3HT, (B) P3MEEMT, and (C) PB2T-TEG as a function of applied voltage (vs. Ag/AgCl). Peak positions as a function of applied voltage (vs. Ag/AgCl) for (D) P3HT, (E) P3MEEMT, and (F) PB2T-TEG. The pink dotted line in F shows the discontinuity in lattice spacing between the neutral and oxidized states of PB2T-TEG. }
\end{figure*}

Figure 1A shows the chemical structure of PB2T-TEG, a conjugated polymer consisting of a repeating backbone unit of two thiophenes and one benzene with two glycolated side chains to facilitate ion uptake and transport. The synthesis and characterization of PB2T-TEG is described in the Supplementary Information, which also includes cyclic voltammetry, absorption spectra, and thermogravimetric analysis of PB2T-TEG (Figure S1 and Figure S2). Derivatives of PB2T-TEG have been used previously as hole extraction materials in lead halide perovskite solar cells.\cite{Zhang2018Modulate} Figure 1B shows cyclic voltammetry (CV) sweeps of PB2T-TEG in 100 mM KCl, indicating that the polymer is indeed electrochemically active in polymer solution, with an oxidation onset of +0.2 V relative to Ag/AgCl (Figure 1B). The spectroelectrochemistry measurements in Figure 1C show a nearly complete bleach by +0.4 V relative to Ag/AgCl, indicating that PB2T-TEG can be rapidly oxidized throughout the entire volume of the polymer with a time constant of 530 $\pm$ 20 ms (Figure S3). Upon increasing the bias in 0.1 V increments from 0 V to +0.7 V, anions compensate for injected holes on the polymer backbone, the $\pi$-$\pi$* absorption peak decreases by approximately 80\% and a redshifted polaron appears (Figure 1D). Volumetric oxidation is also supported by mass changes of the film, as measured by electrochemical quartz crystal microbalance (EQCM) (Figure S4). Based on the electrochemical characterization of PB2T-TEG, we find that it behaves similarly to other conjugated polymers with glycolated side chains in its ability to effectively take up ions from an aqueous solution at a relatively fast rate.\cite{Flagg2019Polymer,Giovannitti2016Controlling,Nielsen2016Molecular}

\begin{figure*}
\includegraphics[width=14cm]{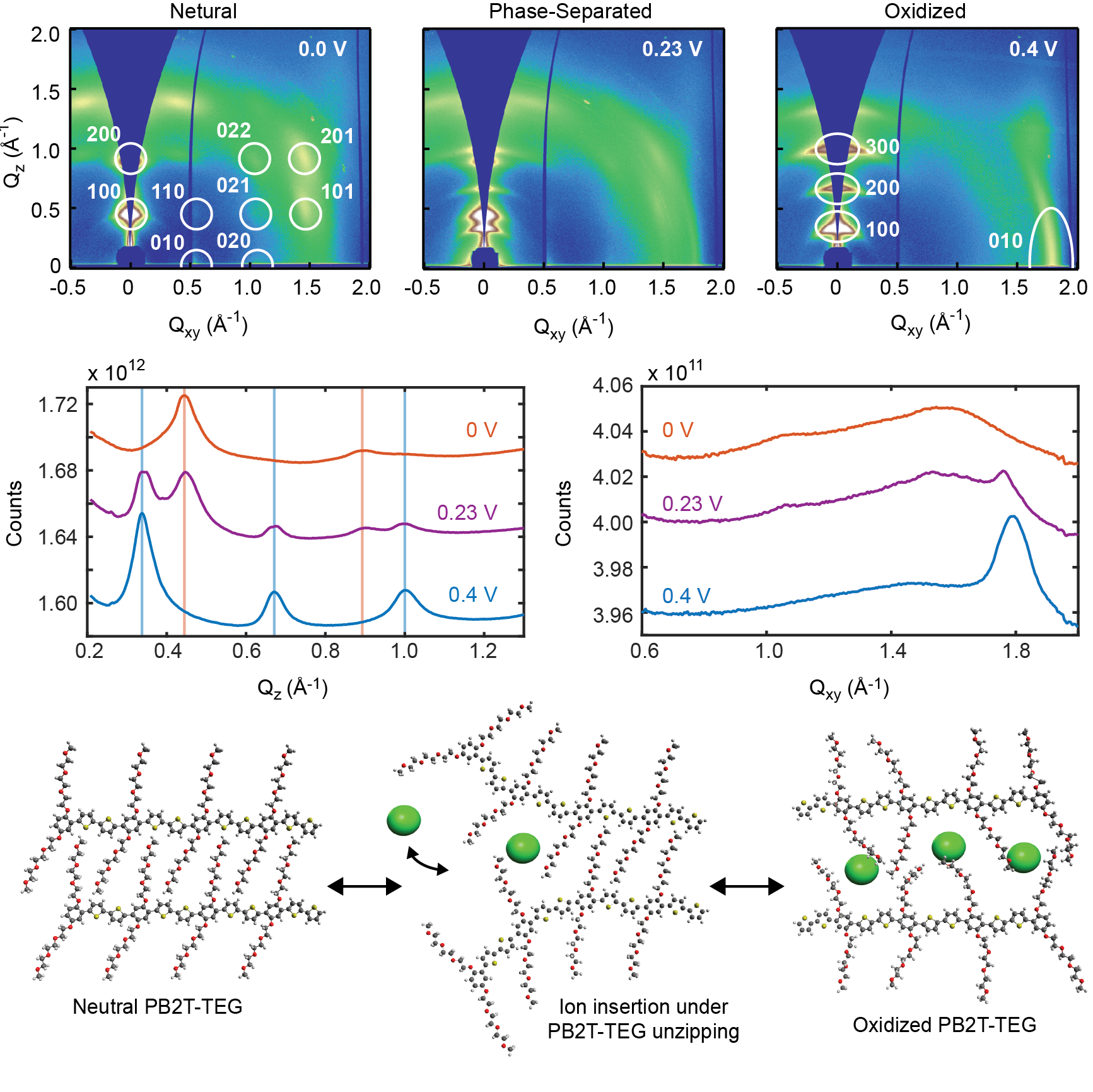}
\caption{\label{fig3} GIWAXS of neutral, phase-separated, and oxidized PB2T-TEG. (A) GIWAXS diffraction patterns of dry PB2T-TEG oxidized, phase-separated, and neutral. Line cuts along the (B) out-of-plane (Q\textsubscript{z}) and (C) in-plane (Q\textsubscript{xy}) directions. (D) Schematic showing the polymer unzipping/zipping mechanism for ion insertion in PB2T-TEG. The green spheres are hydrated Cl\textsuperscript{-} (d = $\sim$6 Å)\cite{Marcus1988Ionic}}
\end{figure*}

We next investigate structural changes in PB2T-TEG as a function of applied bias and compare the observed structural changes to those found in two other conjugated polymers, P3HT and poly(3-{[2-(2-methoxyethoxy)ethoxy]methyl}thiophene-2,5-diyl) (P3MEEMT), that exemplify more typically observed behavior.  Figure 2 compares voltage-dependent X-ray diffraction (XRD) data on a series of polymer films taken at different points in the process of oxidation and ion insertion.  We include P3HT and P3MEEMT as examples of typical conjugated polymer behavior: P3HT has been widely studied as a material for organic transistors,\cite{Bao1996Soluble,Cho2008High-Capacitance} while P3MEEMT is a closely related derivative with the same backbone but possessing ethylene-oxide side chains that exhibits facile redox chemistry and ion uptake in aqueous environments.\cite{Flagg2019Polymer,Dong2019Influence} For these XRD measurements, the polymers are electrochemically oxidized (0 V to +0.7 V vs. Ag/AgCl) in 100 mM KCl (or 100 mM KPF\textsubscript{6} for P3HT), rinsed with deionized water, and dried with N\textsubscript{2} prior to measurement. Figure 2A shows the (100) diffraction peak as a function of applied voltage with the corresponding peak positions shown in Figure 2B. The expansion in the lamellar spacing of P3HT is continuous as a function of applied voltage, expanding by 11\% from 16.6 to 18.5 {\AA} from 0 V to +0.7 V, which is consistent with previous studies using ionic liquids or polymeric ionic liquids as the electrolyte.\cite{Guardado2017Structural,Thomas2018X-Ray} In contrast, the lamellar spacing of P3MEEMT changes minimally by 2\% from 18.8 to 19.2 {\AA} from 0 V to +0.7 V, consistent with previous studies and GIWAXS measurements (Figure S5).\cite{Flagg2019Polymer,Dong2019Influence} In contrast, the lamellar stacking of PB2T-TEG expands discontinuously upon applying a higher potential, expanding by 27\% from 14.3 to 18.1 {\AA} from 0 to +0.7 V. Strikingly, almost all of this expansion occurs over a small window between +0.2 and +0.3 V. This discontinuity in the structural data suggests a fundamentally different mechanism for lattice expansion upon ion insertion for PB2T-TEG, one in which the polymer switches between two structurally-distinct states. Importantly, we note that this discontinuous lamellar expansion is reversible, as the lamellar spacing switches between these two states upon repeated oxidation and reduction (Figure S6).

\begin{figure*}
\includegraphics[width=10cm]{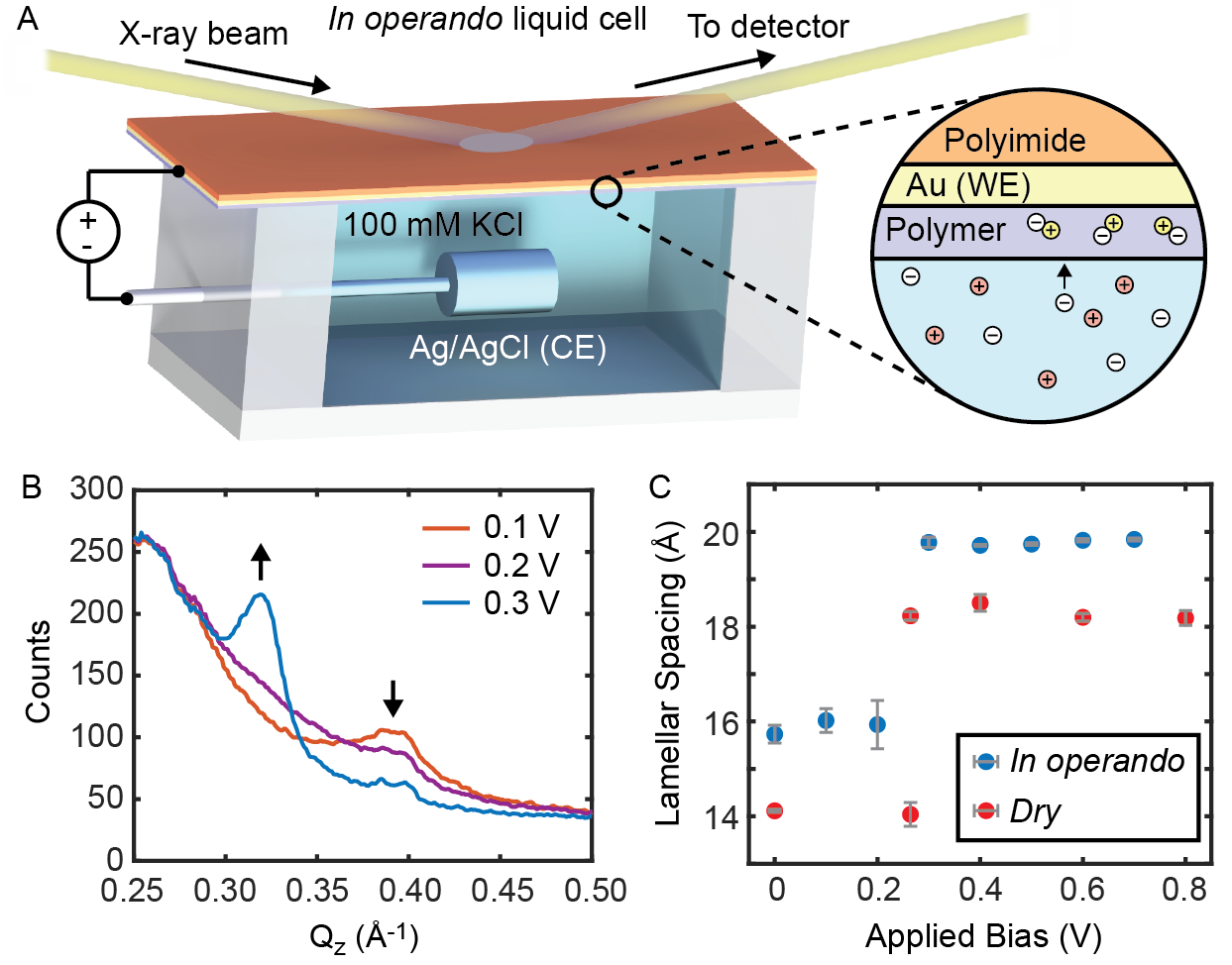}
\caption{\label{fig4}Figure 4: In operando GIWAXS measurements of electrochemical oxidation in PB2T-TEG. (A) Schematic of the custom liquid cell used for in operando GIWAXS measurements. (B) Line cuts of the GIWAXS diffraction pattern along the Q\textsubscript{z} axis for +0.1, +0.2, and +0.3 V. (C) Comparison of the lamellar spacing for the in operando and ex situ GIWAXS measurements, showing a larger lattice spacing for the in operando measurements in both the oxidized and neutral phases.}
\end{figure*}

We further investigate the neutral and oxidized structures of PB2T-TEG using ex situ GIWAXS, focusing on the discontinuity in lattice spacing occurring between +0.2 and +0.3 V (vs. Ag/AgCl).  Figure 3A shows the GIWAXS diffraction patterns of neutral PB2T-TEG, PB2T-TEG oxidized at +0.23 V, and PB2T-TEG oxidized at +0.4 V relative to Ag/AgCl. For the neutral PB2T-TEG, we find that the lamella stacking (100 and 200 peaks) corresponds to a distance equal to one fully extended TEG chain distance (d = ~14.1 Å), suggesting side chain interdigitation.\cite{Schott2015Charge-Transport} The 010 rod corresponds well to the expected distance between the TEG chains along the backbone, or equivalently the length of the repeat unit (d = ~11.3 Å). We propose that the rod at 1.41 Å\textsuperscript{-1} originates from interchain side chain stacking, as this feature corresponds well to typical alkyl stacking distances.\cite{Gruber2015Enabling} While the side chain stacking peaks are present in the XRD data, a $\pi$-$\pi$ stacking peak is absent. Further, we speculate that the two peaks at Q\textsubscript{xy} = 0.38 Å\textsuperscript{-1} (unlabeled in Figure 3A) are the result of twisting in the polymer.\cite{Ko2012Controlled,Carpenter2019Competition,Park2019Tuning} 

Once PB2T-TEG is oxidized at +0.4 V, the out-of-plane (100) spacing significantly increases. At the same time, the glycolated chains lose their crystallinity seen by the disappearance of the h10, h20 and h01 Bragg rods and become less interdigitated as the ions disrupt the side chain interactions. At the same time, the backbones now show evidence of $\pi$-$\pi$ stacking, with a 010 peak at Q\textsubscript{xy} = 1.78 Å\textsuperscript{-1}, suggesting that the backbone untwists (either due to planarization of the quinoidal structure, or due to loss of order from the side chains). GIWAXS spectra acquired at higher bias (+0.6 and +0.8 V) show a similar diffraction pattern as that acquired at +0.4 V (Figure S7). For PB2T-TEG oxidized at +0.23 V, we observe a linear superposition of the GIWAXS patterns acquired at 0 V and +0.4 V. This superposition is also clear from line-cuts taken from the in-plane and out-of-plane directions of the GIWAXS patterns in Figure 3C and Figure 3D. This linear superposition of the oxidized (+0.4 V) and neutral states (0 V) suggests that the polymer is composed of a phase-separated blend of ion-rich and ion-poor structural phases. 

Overall, in the neutral form, the polymer structure is driven by side chain crystallization (perhaps because the backbone does not support strong $\pi$-$\pi$ stacking), while in the oxidized/ion compensated form, the polymer crystal structure is driven by $\pi$-$\pi$ stacking, with the side chains being more disordered. Next, we discuss a possible mechanism to account for the observed structural changes.  We suggest that PB2T-TEG undergoes an unzipping/zipping mechanism occurring upon oxidation and reduction, as shown in Figure 3D. Initially in its neutral structural phase, PB2T-TEG has interdigitated, side chains and no significant $\pi$-$\pi$ stacking. Following oxidation of the backbone under an applied bias, the glycolated side chains lose their crystallinity, and the backbone (which becomes more quinoidal upon oxidation) forms $\pi$-$\pi$ stacking interactions through what we hypothesize is an unzipping and potentially untwisting motion of the polymer. Such an unzipping mechanism could also underpin the origin of the phase-separated state. We propose that it should be energetically-favorable to electrochemically oxidize regions of the polymer that are already unzipped, presumably leading to nucleation and growth-like kinetics of ion insertion. Temperature-dependent XRD measurements on the phase-separated film show that the two phases mix with increasing temperature and then demix as the temperature is lowered (Figure S8). This observation is consistent with the proposed unzipping mechanism, as the side chains most likely melt at higher temperature, leading to reversible mixing of previously phase-separated regions of the polymer. 

Finally, we show that the structural phase transition persists under working electrochemical conditions using in operando GIWAXS.  Figure 4A shows the in operando GIWAXS electrochemical cell for interrogating changes in polymer structure upon electrochemical oxidation in an aqueous electrolyte environment (with additional images of the liquid cell shown in Figure S9). The cell consists of an Au-coated polyimide window with the conjugated polymer layer located between the Au layer and the aqueous electrolyte (100 mM KCl). A Ag/AgCl pellet electrode is used as the counter electrode, which is placed in a well filled with aqueous electrolyte and surrounded by a PDMS spacer. The X-ray beam penetrates the Au/polyimide window to probe changes in the polymer structure upon applying an electrochemical bias. Figure 4B shows line cuts of the GIWAXS diffraction pattern in the lamellar direction (Q\textsubscript{z}). The line cuts show the disappearance of the peak corresponding to the neutral state of PB2T-TEG and appearance of the peak correspond to its oxidized state (the full GIWAXS patterns are shown in Figure S10). Upon returning to 0 V, the peak corresponding to neutral PB2T-TEG reappears, indicating that this process is also reversible in operando (Figure S11). In Figure 4C, we compare the lattice spacing in the lamellar direction (Q\textsubscript{z}) for ex situ and in operando GIWAXS characterization. We observe the same discontinuity between +0.2 and +0.3 V with the in operando measurements, yet the lattice spacing for the in operando measurements with electrolyte present is larger for both oxidized and neutral samples by approximately 12\% and 8\%, respectively. This observation is consistent with previous reports showing that polymer lattice spacing can increase in a hydrated environment. Importantly, the in operando GIWAXS measurements demonstrate that the structural phase transition occurs in a working device-relevant environment. 

\begin{figure*}
\includegraphics[width=14cm]{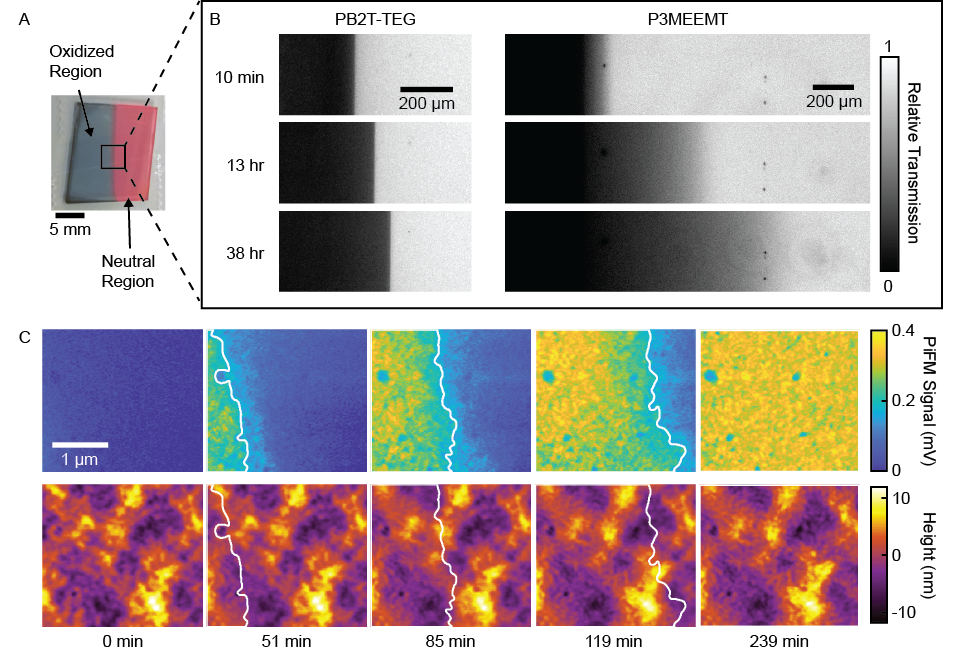}
\caption{\label{fig5} Ion front movement in PB2T-TEG and P3MEEMT. (A) Example image of the samples used for ion front imaging. The gray area is oxidized in 100 mM KCl at +0.7 V (vs. Ag/AgCl) and the red area is neutral. (B) Optical transmission images of the interface between oxidized and neutral regions of the film for PB2T-TEG and P3MEEMT. The transmission images are acquired in the same spectral region as the polaron absorption (750 $\pm$ 25 nm). (C) Time series of spatially correlated PiFM and topography images as the interface migrates in PB2T-TEG. The white line designates the 50\% intensity threshold from the PiFM channel. }
\end{figure*}

The structural phase transition has significant implications on how ions move through the crystalline polymer matrix. To interrogate how the electrochemically-induced phase transition impacts ion motion, we compare ion transport in PB2T-TEG and P3MEEMT. We choose to compare PB2T-TEG to P3MEEMT because P3MEEMT shows minimal changes in lattice spacing upon ion injection. To directly visualize ion transport, we use moving front diffusion experiments. Unlike previous moving front experiments,\cite{Aoki1992Photographic,Wang2004Visualizing,Johansson2004Moving,Stavrinidou2013Direct} we do not apply an electrochemical bias in situ. Instead, we electrochemically oxidize half the sample and watch the interface relax to equilibrium through ion diffusion under dry conditions without an applied bias. To create an interface between oxidized and neutral regions of the polymers, we electrochemically oxidize half of a polymer sample (+0.7 V vs. Ag/AgCl) by submerging half of the sample in an aqueous electrolyte (100 mM KCl), applying an electrochemical bias (+0.7 V vs. Ag/AgCl), and then removing the sample after 30 s, resulting in a sample that is half oxidized and half neutral (Figure 5A). We then watch the interface relax to equilibrium using optical transmission imaging of the polaron absorption peak (at 750 $\pm$ 25 nm) as ions diffuse. Figure 2B and Movie S1 shows the interface motion for PB2T-TEG and P3MEEMT (with line profiles shown in Figure S12). The P3MEEMT interface between oxidized and neutral regions broadens and becomes less sharp out with time, with a 50\% transmission threshold at 720 $\mu$m after 38 hrs, as expected for Fickian diffusion of the ions. In contrast, the interface between the oxidized and neutral regions in PB2T-TEG remains sharp with time ($<$ 1 $\mu$m), moving 140 $\mu$m at 38 hrs. This interface remains sharp for as long as six months after electrochemical oxidation (Figure S13). 

To confirm that the ion front remains sharp, even below the diffraction limit, we also tracked ion motion as a function of time in PB2T-TEG at the nanoscale using super-resolution photoinduced force microscopy (PiFM).  In doing so, we find that the ion front is also sharp at the nanoscale as the interface moves (Figure 5D). For imaging ion motion, we electrochemically oxidize the half of the film from a 100 mM KPF\textsubscript{6} solution (Figure S14), so that we can use the P-F IR active vibrational mode of PF\textsubscript{6}\textsuperscript{-} to detect the presence of ions within the PB2T-TEG film (Figure S15). We capture simultaneous images of the infrared signal of PF\textsubscript{6}\textsuperscript{-} and the sample topography as the interface between the oxidized and neutral regions of the polymer transits the field-of-view (Figure 5D, Figure S16, Movie S3, and Movie S4). We find that the ion front remains sharp (~700 nm 90/10 width of a line-cut) as the phase front propagates. Along with changes in the PiFM signal, we also observe significant changes in the film topography as ions move through the materials (Figure S17). The film becomes significantly more wrinkled in regions of the polymer that contain ions, presumably due to the lattice expansion of the film.

While this non-Fickian transport appears startling at first, we hypothesize that the polymer phase change, rather than the intrinsic diffusivity of the ions, is what limits ion motion and hence maintains a sharp interface between neutral and oxidized regions of the film. This observation is reminiscent of the sharp propagating front observed upon solvent infiltration into some glassy polymers, which is often called Case II diffusion and results from the mechanical response of the polymer limiting diffusion.\cite{thomas_theory_1982} This hypothesis is also consistent with numerical calculations of concentration-dependent diffusion, which show a sharp moving front can be preserved when the diffusion constant of the ions in the oxidized and neutral region differs by at least three orders of magnitude (Figure S18 with additional details in the Supplementary Information). We note that similar sharp interfaces have previously been observed in inorganic materials for energy storage, such as in interfaces between lithium-rich and lithium-poor iron phosphate (LiFePO\textsubscript{4} / FePO\textsubscript{4}),\cite{Lim2016Origin} but not in organic materials.

	We have demonstrated that conjugated polymers can undergo reversible structural phase transitions upon electrochemical insertion of ions, in analogy with inorganic materials like LiFePO\textsubscript{4}. We suggest that the structural phase transition occurs due to melting of the glycolated side-chains upon ion insertion, accompanied by the formation $\pi$-$\pi$ stacking interactions of the polymer backbone in the oxidized state. The kinetics of the phase transition can be described by an unzipping motion of the polymer backbone.  At intermediate biases, the films exhibit phase-separation between ion-rich and ion-poor regions of the polymer film. Using a custom liquid cell for in operando GIWAXS measurements, we also show that the phase transition occurs in aqueous electrolytes. In addition to its fundamental interest as a structural motif, this phase transition has important implications for the mixed-conduction behavior of conjugated polymers: whereas the polymers P3MEEMT that does not exhibit a phase transition exhibits normal Fickian diffusion, ions in PB2T-TEG propagate as a sharp front, that does not spread out in time over periods of days to months. We speculate that this behavior could be used to engineer desirable electrochemical properties, perhaps aiding retention of electrochemical states in computing applications,\cite{Burgt2017non-volatile,Fuller2019Parallel,Burgt2018Organic} or exploited to improve energy storage in conjugated polymers, where structural phase transitions might enable flatter voltage profiles during discharge due to phase coexistence between the oxidized and reduced phases. As a result, engineering structural phase transitions in conjugated polymers could become an important new design principle.

\section*{Acknowledgments}

L. Q. F and R. J. Q.’s contributions are based in part on work supported by the expiring National Science Foundation award, NSF DMR-1607242, and subsequent support from NWIMPACT SEED award funding. C. G. B. is a Washington Research Foundation Postdoctoral Fellow. CHANG-ZHI STUDENT and C.-Z. L. thank the support from National Natural Science Foundation of China (Nos. 21722404 and 21674093) for supporting the synthesis of PB2T-TEG.  The photoinduced force microscopy (PiFM) and X-ray diffraction (XRD) were conducted at the Molecular Analysis Facility, a National Nanotechnology Coordinated Infrastructure site at the University of Washington that is supported in part by the National Science Foundation (grant ECC-1542101), the University of Washington, the Molecular Engineering \& Sciences Institute, the Clean Energy Institute and the National Institutes of Health. This research used beamline 7.3.3 of the Advanced Light Source (ALS), which is a DOE Office of Science User Facility under contract no. DE-AC02-05CH11231. The authors thank C. Zhu and E. Schaible at the ALS for assistance with GIWAXS data acquisition and analysis. 

\section*{Author Contributions}

C.G.B., L.Q.F. and D.S.G. conceived of the experiments and wrote the manuscript. C.-Z.L., K.Y., and T.R. synthesized and characterized PB2T-TEG. J.W.O. and C.K.L. synthesized and characterized P3MEEMT. C.G.B. collected and interpreted electrochemical, spectroelectrochemistry, XRD, optical transmission, and PiFM data. C.G.B., L.Q.F., and R.J.Q. collected E-QCM and GIWAXS data and built the in operando liquid cell. Y.D., C.K.L., C.G.B. and D.W.D. interpreted the GIWAXS data. C.G.B. performed diffusion calculations. 
\section*{References}

\bibliography{bibtexrefs.bib}

\end{document}